\documentclass[epj]{svjour}
\usepackage{latexsym}
\usepackage{graphics}
\usepackage{graphicx}
\usepackage{float}
\usepackage{ragged2e}
\usepackage{dcolumn}
\usepackage{hyperref}

\usepackage{adjustbox}
\usepackage{xcolor}
\usepackage{dirtytalk}
\usepackage{cite}
\usepackage{svg}
\begin{document}
\title{Study of quasi-projectile properties at Fermi energies in $^{48}$Ca projectile systems}
\subtitle{(The FAZIA Collaboration)}
\author{S. Upadhyaya\inst{1}\thanks{email: sahil.upadhyaya@ifj.edu.pl}\thanks{Current address: Institute of Nuclear Physics Polish Academy of Sciences (IFJ-PAN), 31-342 Krakow, Poland}
         \and K. Mazurek\inst{2}
         \and T. Kozik\inst{1}
         \and D. Gruyer\inst{3}
         \and G. Casini\inst{4}
         \and S. Piantelli\inst{4}
         \and L. Baldesi\inst{5,4}
         \and S. Barlini\inst{4,5}
         \and B. Borderie\inst{6}
         \and R. Bougault\inst{3}
         \and A. Camaiani\inst{4,5}
         \and C. Ciampi\inst{7}
         \and M. Cicerchia\inst{8,9}
         \and M. Ciemała\inst{2}
         \and D. Dell’Aquila\inst{10,11}
         \and J. A. Dueñas\inst{12}
         \and Q. Fable\inst{13}
         \and J. D. Frankland\inst{7}
         \and F. Gramegna\inst{14}
         \and M. Henri\inst{15}
         \and B. Hong\inst{16,17}
         \and A. Kordyasz\inst{18}
         \and M. J. Kweon\inst{19}
         \and N. Le Neindre\inst{3}
         \and I. Lombardo\inst{20,21}
         \and O. Lopez\inst{3}
         \and T. Marchi\inst{14}
         \and S. H. Nam\inst{16,17}
         \and J. Park\inst{16,17}
         \and M. Pârlog\inst{3,22}
         \and G. Pasquali\inst{4,5}
         \and S. Valdré\inst{4}
         \and G. Verde\inst{20,13}
         \and E. Vient\inst{3}
         \and M. Vigilante\inst{23}
%
}                     
%
%
\institute{Marian Smoluchowski Institute of Physics, Jagiellonian University, 30-348 Krakow, Poland
           \and
           Institute of Nuclear Physics Polish Academy of Sciences (IFJ-PAN), 31-342 Krakow, Poland
           \and
           Université de Caen Normandie, ENSICAEN, CNRS/IN2P3, LPC Caen UMR6534, F-14000 Caen, France 
           \and
           INFN - Sezione di Firenze, 50019 Sesto Fiorentino, Italy
           \and
           Dipartimento di Fisica e Astronomia, Università di Firenze, 50019 Sesto Fiorentino, Italy
           \and
           Université Paris-Saclay, CNRS/IN2P3, IJCLab, F-91405 Orsay, France
           \and
           Grand Accélérateur National d’Ions Lourds (GANIL), CEA/DRF–CNRS/IN2P3, Boulevard Henri Becquerel, F-14076 Caen, France
           \and
           Dipartimento di Fisica e Astronomia, University of Padova, Padova, Italy
           \and
           INFN - Sezione di Padova, 35121 Padova, Italy
           \and
           Dipartimento di Fisica "Ettore Pancini", University of Naples "Federico II", Naples, Italy
           \and
           INFN-Sezione di Napoli, Naples, Italy
           \and
           Departamento de Ingeniería Eléctrica y Centro de Estudios Avanzados en Física, Matemáticas y Computación, Universidad de Huelva, 21007 Huelva, Spain
           \and
           Laboratoire des 2 Infinis - Toulouse (L2IT-IN2P3), Université de Toulouse, CNRS, UPS, F-31062 Toulouse Cedex 9, France
           \and
           INFN - Laboratori Nazionali di Legnaro, 35020 Legnaro, Italy
           \and
           CEA/INSTN/UECC, Les Vindits - 143, chemin de la Crespinière 50130 Cherbourg-Octeville, France
           \and
           Center for Extreme Nuclear Matters (CENuM), Korea University, Seoul 02841, Republic of Korea
           \and
           Department of Physics, Korea University, Seoul 02841, Republic of Korea
           \and
           Heavy Ion Laboratory, University of Warsaw, 02-093 Warszawa, Poland
           \and
           Department of Physics, Inha University, Incheon 22212, Republic of Korea
           \and
           INFN - Sezione di Catania, 95123 Catania, Italy
           \and
           Dipartimento di Fisica e Astronomia, Università di Catania, via S. Sofia 64, 95123 Catania, Italy
           \and
           National Institute for Physics and Nuclear Engineering, RO-077125 Bucharest-Măgurele, Romania
           \and
           INFN - Sezione di Napoli, Via Cintia, I-80126 Napoli, Italy.
           }         
\date{Received: date / Revised version: date}
%
\abstract{
The emission of the pre-equilibrium particles during nuclear collisions at moderate beam energies is still an open question. This influences the properties of the compound nucleus but also changes the interpretation of the quasi-fission process. A systematic analysis of the data obtained by the FAZIA collaboration during a recent experiment with a neutron rich projectile is presented. The full range of charged particles detected in the experiment is within the limit of isotopic resolution of the FAZIA detector. Quasi-projectile ($QP$) fragments were detected in majority thanks to the forward angular acceptance of the experimental setup which was confirmed by introducing cuts based on the HIPSE event generator calculations. The main goal was to compare the experimental results with the HIPSE simulations after introducing these cuts to investigate the influence of the n-rich entrance channel on the $QP$ fragment properties. More specifically, the lowering of $N/Z$ of $QP$ fragments with beam energy was found to be present since the initial phase of the reaction. Thus, pre-equilibrium emissions might be a possible candidate to explain such an effect.
\PACS{
      {PACS-key}{discribing text of that key}
     } 
} 
\maketitle
\section{Introduction}
\label{intro}
Nuclear reactions in the intermediate energy range (20--100 MeV/u) have been of interest in studying various nuclear properties with respect to the beam energy, mass of the system and centrality of the collision. Excitation energies of the hot nuclei in these reactions can be close to or even higher than their total binding energies. The production of a large range of fragments is one of the main features of such reactions, as a consequence of the various processes that are responsible for fragment emission, spread over different time scales.

From an experimental point of view, the most important factor in studying such reactions is a high capability of identifying the charge ($Z$) and mass ($A$) of each emitted fragment. Among the many particle identification devices \cite{multics,garfield,chimera,kratta,nimrod,hyra}, the one designed and developed by the FAZIA collaboration \cite{fazia} represents the state of the art for this kind of studies at Fermi energies. At present, the FAZIA detector is able to detect the full range of fragment charge with a mass resolution up to $Z\sim25$ in the typical phase-space region of quasi-projectile fragments at Fermi energies. This was achieved mainly by using silicon detectors with high dopant homogeneity obtained with the n-TD process and specific crystallographic orientation, their reverse mounting so that the particles enter from the low-field side, usage of dedicated pre-amplifiers, located as close as possible to the detector and extensive optimization of the digital treatment of the sampled pulse shapes of both charge and current signals \cite{Bougault2014,PASTORE201742}.

With the excellent isotopic resolution of FAZIA, numerous experimental endeavors have been devoted to the investigation of isospin (neutron-to-proton ratio -- $N/Z$) related physics (for example, see \cite{piantelli2023,PhysRevC.87.054607,PhysRevC.102.044607,PhysRevC.101.034613,PhysRevC.103.014603,PhysRevC.103.014605,PhysRevC.106.024603}). This was mainly realized by using projectile-target combinations with different $N/Z$ compositions. The isospin degree of freedom and its influence on the reaction dynamics and on the subsequent decay processes has been widely studied by various groups (see \cite{Pagano2020,epja30.2006,PhysRevC.79.064614,PhysRevC.76.024606,LI2008113,Di_Toro_2010} and references therein). The experimental observables associated with the isospin content of the reaction products can be used to extract information on the symmetry energy ($E_{sym}$) term of the nuclear equation of state ($nEoS$), via comparison with theoretical models \cite{Pagano2020,epja30.2006,PhysRevC.76.024606,LI2008113,Di_Toro_2010,PhysRevLett.84.1120,PhysRevC.82.014608,GERACI2004173,SOULIOTIS200435,PhysRevLett.92.062701,PhysRevLett.102.122701,PhysRevC.82.051603,PhysRevC.62.041605,PhysRevC.86.014610,BARAN2005335,PhysRevLett.94.032701,PhysRevC.76.034603,23,PhysRevC.68.024605,COLONNA2008454c,PhysRevC.80.014322,PhysRevC.79.064615,fable2023study} and study isospin transport (differential flux of neutrons vs protons and nucleon density) in dissipative collisions at low energies, e.g. \cite{PhysRevC.37.1783,PhysRevC.38.195} and at Fermi energies \cite{PhysRevC.79.064614,PhysRevLett.92.062701,PhysRevLett.102.122701,PhysRevC.82.051603,PhysRevC.62.041605,PhysRevC.86.014610,PhysRevC.76.034603,PhysRevC.74.051602,PhysRevC.81.034603,PhysRevC.86.021603,PhysRevC.106.024605,Ciampi_2023,PhysRevC.107.014604,PhysRevC.108.054611}.

In this paper we show a systematic analysis of data obtained in a recent FAZIA experiment by bombarding with a neutron rich $^{48}$Ca projectile at 25 MeV/u and 40 MeV/u on $^{12}$C, $^{27}$Al and $^{40}$Ca targets. The experiment was based on an approach to study the effect of a neutron rich projectile on the properties of reaction products mainly for semi-peripheral/peripheral collisions. These collisions are basically binary dissipative events which lead to the production of two primary main fragments, a quasi-projectile ($QP$) and a quasi-target ($QT$). The experimental data are also compared with the simulations produced by the HIPSE (Heavy-Ion Phase Space Exploration) event generator \cite{PhysRevC.69.054604,lacroix:in2p3-00023917}.

The initial conditions of neutron richness can be diminished due to pre-equilibrium (fast) emissions from the projectile. These pre-equilibrium emissions were reported for neutrons and light particles within the Fermi energy range with pre-equilibrium neutron multiplicity increasing with the beam energy \cite{PhysRevC.55.1900}. For our reactions induced by a n-rich projectile, if there is an even slight larger probability for fast neutron emission versus charged emissions (e.g. more free neutrons than protons) this will result in a net loss in the neutron richness of the remaining system. Our goal here is to ultimately check the effect of this diminished neutron richness in the first phases of the reaction on the $QP$ fragment with the help of HIPSE. However, there is an important point to be mentioned; as also clearly discussed in an our recent paper \cite{piantelli2023} dealing with similar systems, in the case of the light $^{12}$C target, the description based on the binary channel scenario weakens due to the strong size asymmetry of the entrance channel. In this case, apart from very grazing collisions, the reactions mostly form a single excited source (a kind of composite system) more than a real pair $QP-QT$. However, since we are ultimately interested in the events more ascribable to semi-peripheral reactions, the heavier fragment in the events is of our main concern, although we are aware that the association of this fragment as the $QP$ is more valid for the reaction on $^{27}$Al or $^{40}$Ca than on $^{12}$C. On the other hand, the data for $^{40}$Ca target system is present only for 25 MeV/u beam energy and cannot be used to check beam energy dependence on fragment properties. With these \say{caveats} in mind, in the initial part of the paper we decide to show the results for all systems for a clearer systematic and comparative analysis.

This article is organised in such a way that in Section II the details of the experimental setup along with a discussion on the obtained experimental data are given. The information about the HIPSE event generator and simulated data filtering is given in Section III, and Section IV discusses the selection conditions introduced according to the HIPSE predictions on the QP fragments and shows further comparison between the experimental data and simulations. Summary and conclusion are given in the Section V. 

\section{\label{experiment}Experimental setup and data}

\begin{figure}[!b]
\includegraphics[width=1.0\columnwidth]{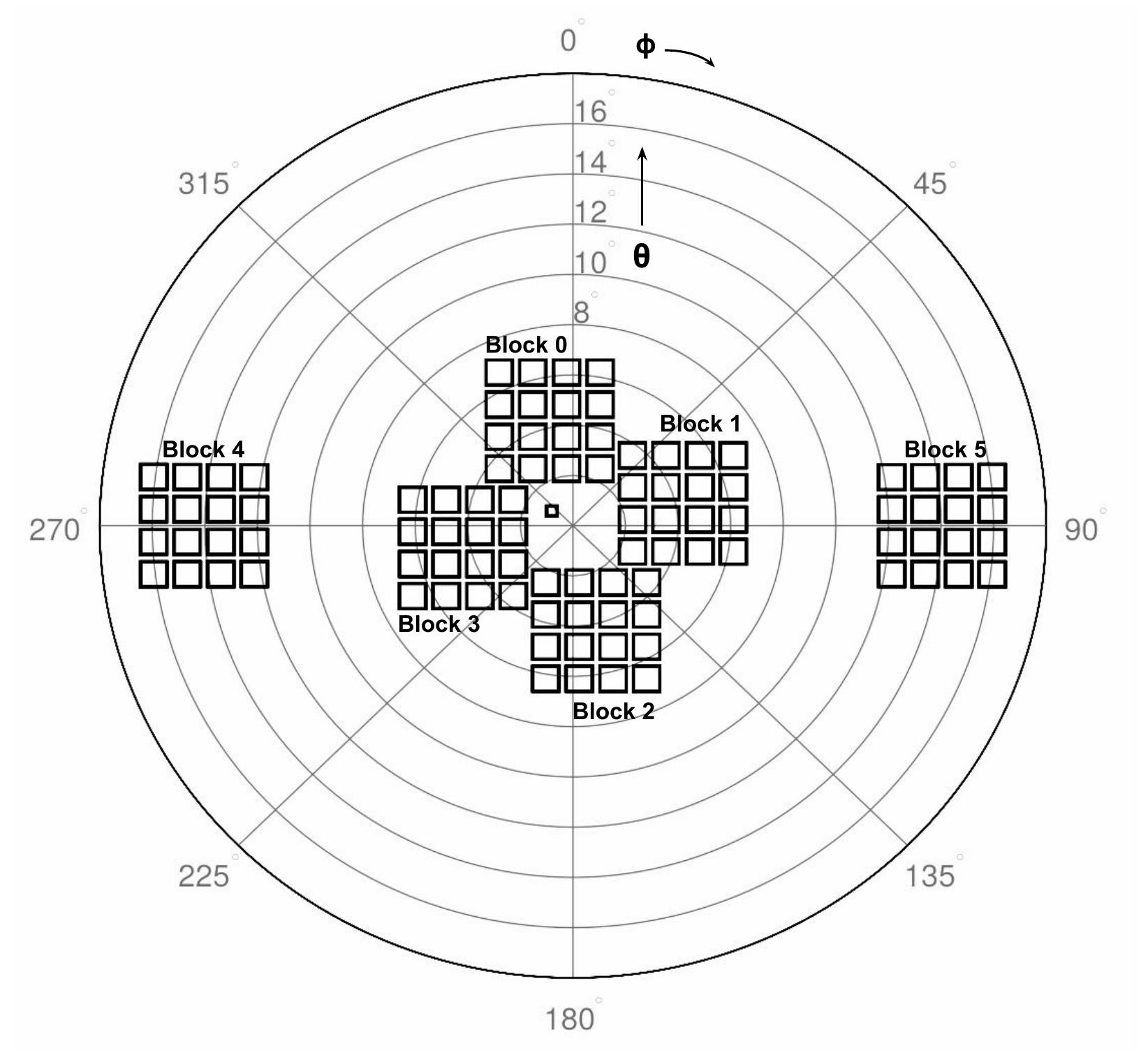}
\caption{Polar representation of the detector geometry of FAZIA-PRE experimental setup with the corresponding angular coverage. The four blocks in wall configuration cover $\sim$1.7--7.6$^{\circ}$ of $\theta$. The remaining 2 blocks on the sides have a coverage of $\sim$11.5--16.7$^{\circ}$ of $\theta$. Directions of used polar coordinates, $\theta$ and $\phi$ are marked.}
\label{1}
\end{figure}

The FAZIA-PRE experiment (also discussed in \cite{piantelli2023}) was performed at the Laboratori Nazionali del Sud (LNS-INFN), Catania, Italy using 6 FAZIA blocks (Fig. \ref{1}) placed 1 m far from the target inside the LNS-Ciclope chamber. Therefore, a total of 96 (6x16) Si (300 $\mu$m)$-$Si (300/500 $\mu$m)$-$CsI (10 cm) telescopes were used. The overall angular coverage of the detector setup was in the range $\theta\sim$1.7--7.6$^{\circ}$ and $\sim$11.5--16.7$^{\circ}$ as shown in Fig. \ref{1}.

\begin{table}[!b]
\caption{FAZIA-PRE experimental details. $^{48}_{20}$Ca projectile on $^{12}_{6}$C, $^{27}_{13}$Al and $^{40}_{20}$Ca targets at 25 and 40 MeV/u beam energies ($E_{B}$) along with their corresponding target thicknesses ($t$), beam velocities ($v_{B}$), centre-of-mass velocities ($v_{CM}$), energies in centre-of-mass ($E_{CM}$), total isospin of the system ($N/Z_{proj+tar}$) and grazing angles in laboratory frame ($\theta_{gr}$).}
\label{tab1}
\centering{
\begin{tabular}{c|c|c|c|c|c}
\hline
Projectile                                     & \multicolumn{5}{c}{$^{48}_{20}$Ca}                                                                 \\ \hline
$E_{B}$ {[}MeV/u{]}              & \multicolumn{3}{c|}{25}                                    & \multicolumn{2}{c}{40}                \\ \hline
Target                                         & $^{12}_{6}$C       & $^{27}_{13}$Al    & $^{40}_{20}$Ca    & $^{12}_{6}$C       & $^{27}_{13}$Al    \\ \hline
$t$ {[}$\mu$g/cm$^{2}${]}         & 239                & 216               & 500               & 239                & 216               \\ \hline
$v_{B}$ {[}cm/ns{]} & 6.8             & 6.8           & 6.8           & 8.5             & 8.5           \\ \hline
$v_{CM}$ {[}cm/ns{]} & 5.5             & 4.4           & 3.8           & 7.0             & 5.5           \\ \hline
$E_{CM}$ {[}MeV/u{]}             & 4.0            & 5.7            & 6.2           & 6.4            & 9.2            \\ \hline
$N/Z_{proj+tar}$                      & 1.31 & 1.27 & 1.2 & 1.31 & 1.27 \\ \hline
$\theta_{gr}$                      & 0.9$^{\circ}$ & 1.8$^{\circ}$ & 2.7$^{\circ}$ & 0.5$^{\circ}$ & 1.1$^{\circ}$
\\ \hline
\end{tabular}}
\end{table}

In this experiment, $^{48}$Ca beams bombarded $^{12}$C, $^{27}$Al and $^{40}$Ca targets at 25 MeV/u and, $^{12}$C and $^{27}$Al targets at 40 MeV/u. As said, the motivation for this choice of reaction systems is to study the fragment properties in the presence of a initial neutron richness. The $^{48}$Ca projectile is a neutron rich nucleus with $N/Z=1.4$. The targets $^{12}$C and $^{40}$Ca are symmetric ($N/Z=1$) and $^{27}$Al has one extra neutron making its $N/Z=1.07$. The $N/Z$ of the whole system ($N/Z_{proj+tar}$) ranges from 1.2 to 1.31. The values are shown in the Table \ref{tab1} along with other experimental details like their corresponding target thicknesses ($t$), beam velocities ($v_{B}$), centre-of-mass velocities ($v_{CM}$), energies in centre-of-mass ($E_{CM}$), total isospin of the system ($N/Z_{proj+tar}$) and grazing angles in laboratory frame ($\theta_{gr}$).

\begin{figure}[!b]
\centering{
\includegraphics[width=0.85\columnwidth]{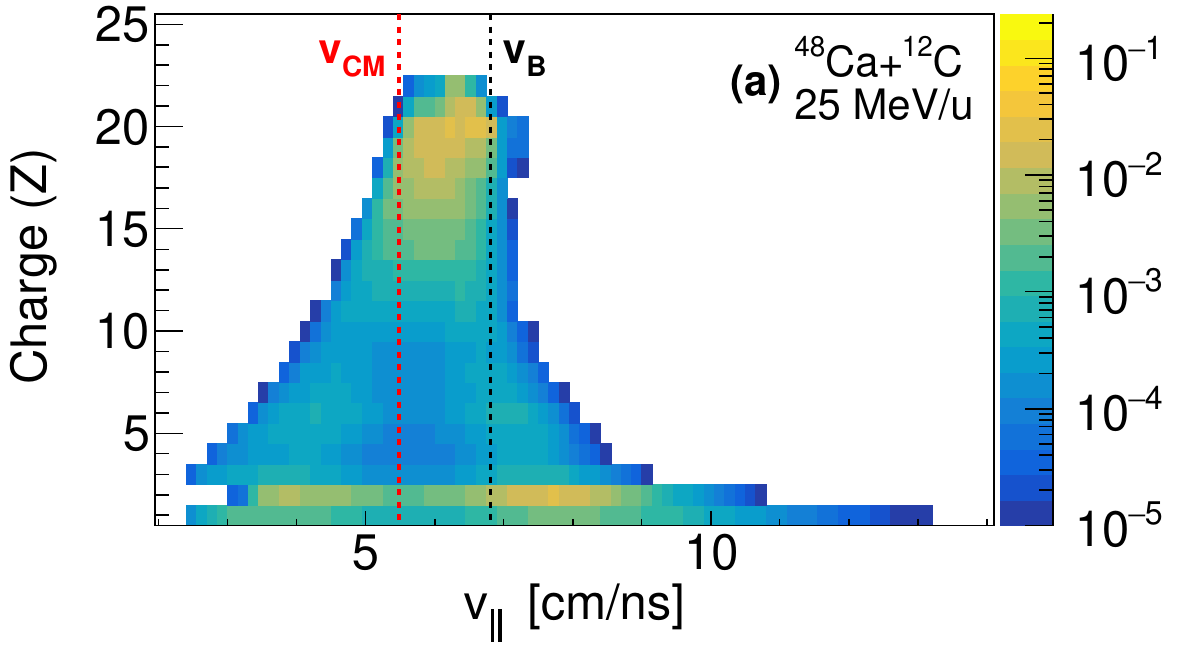}
\includegraphics[width=0.85\columnwidth]{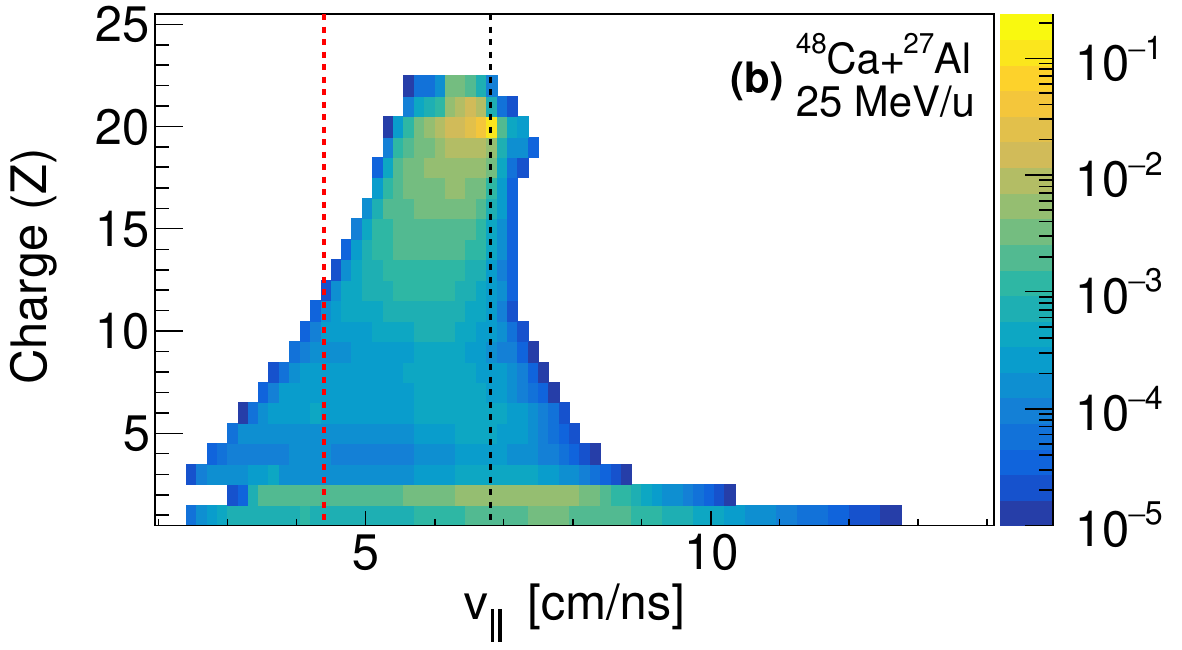}
\includegraphics[width=0.85\columnwidth]{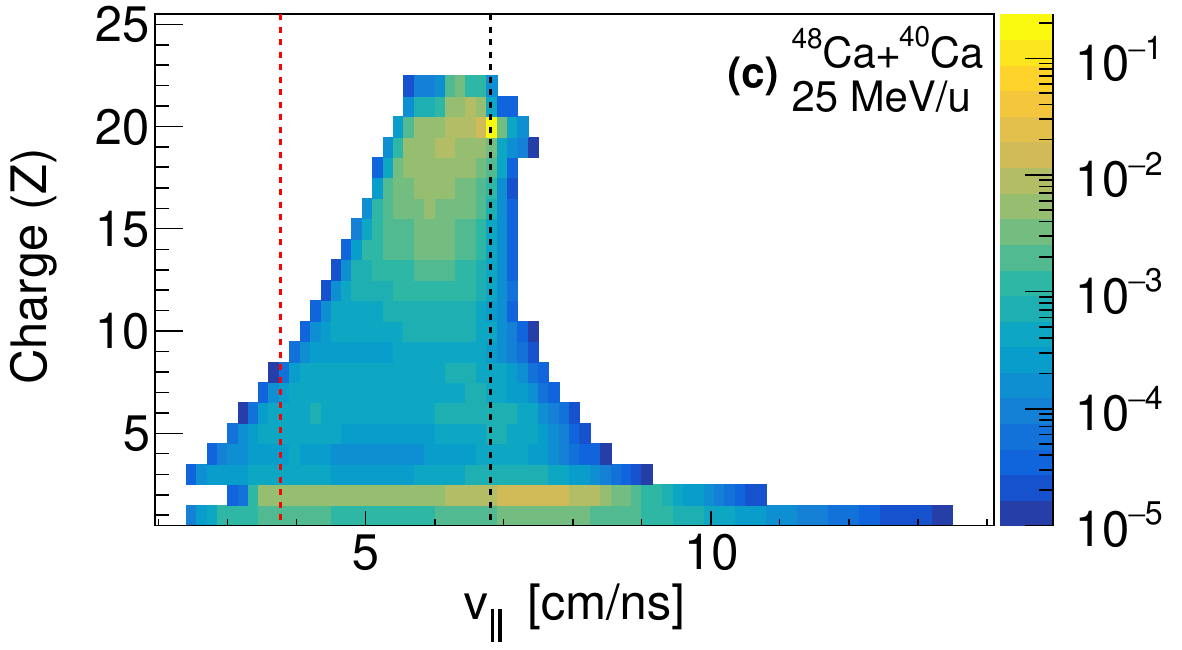}
\includegraphics[width=0.85\columnwidth]{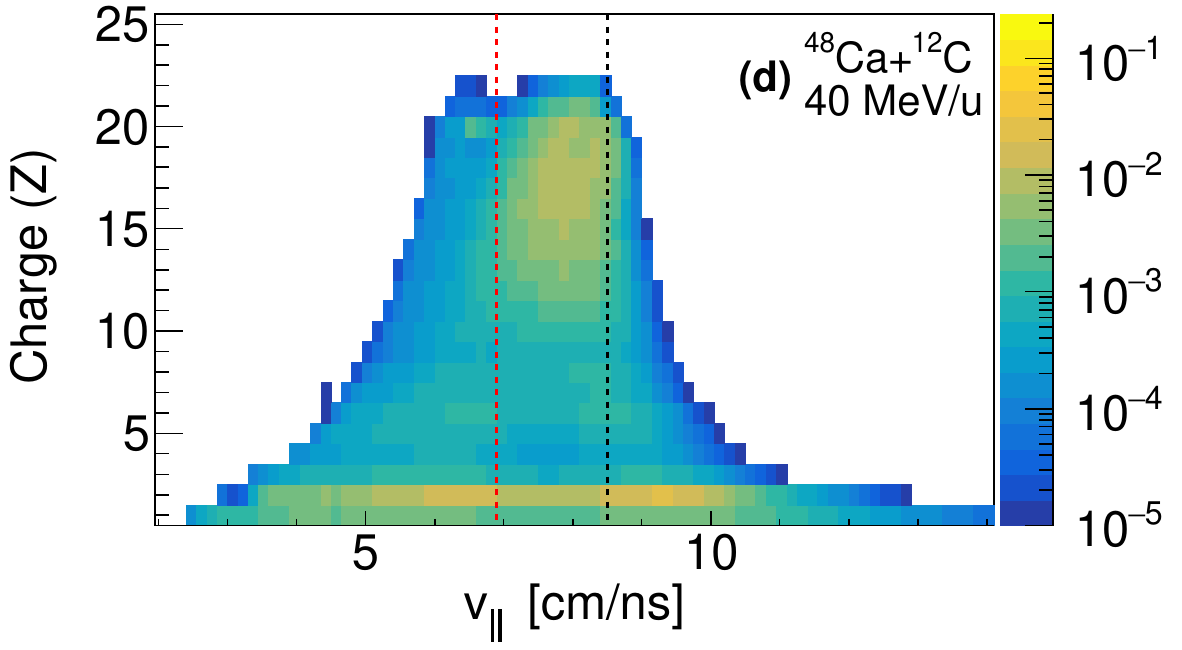}
\includegraphics[width=0.85\columnwidth]{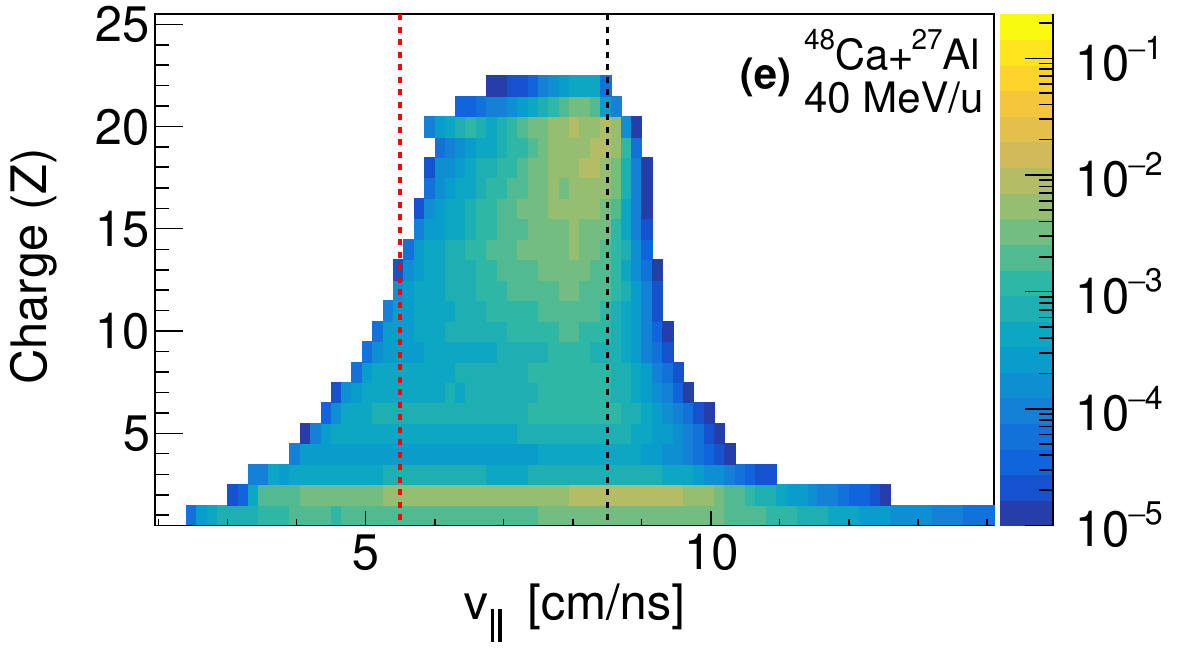}}
\caption{Experimental data: Fragment charge ($Z$) vs parallel velocity ($v_{\parallel}$) for all fragments detected in the FAZIA-PRE experiment. All systems are shown. Vertical red and black dashed lines show the corresponding centre-of-mass velocities ($v_{CM}$) and beam velocities ($v_{B}$), respectively.}
\label{zvz}
\end{figure}

The very forward angular acceptance of the experimental setup allows to efficiently detect  $QP$ fragments. The $QT$ fragments cannot be detected because they are mainly spread at relatively large polar angles and cannot reach the detector at such forward angles. This fact can be well observed from the Fig. \ref{zvz} which shows the charge $Z$ and parallel velocity ($v_{\parallel}$) correlation in the laboratory frame for all the detected fragments. As expected, for all the five systems under consideration, the $Z$ vs $v_{\parallel}$ correlation presents an intense region around the projectile $Z$ (= 20) and the corresponding beam velocities ($v_{\parallel}$ = $v_{B}$) marked by vertical black dashed lines. This intense region in the data corresponds to the $QP$ fragments. The centre-of-mass velocity ($v_{\parallel}$ = $v_{CM}$) is also marked for each system by vertical red dashed lines.

\begin{figure}[!bt]
\centering{\includegraphics[width=0.8\columnwidth]{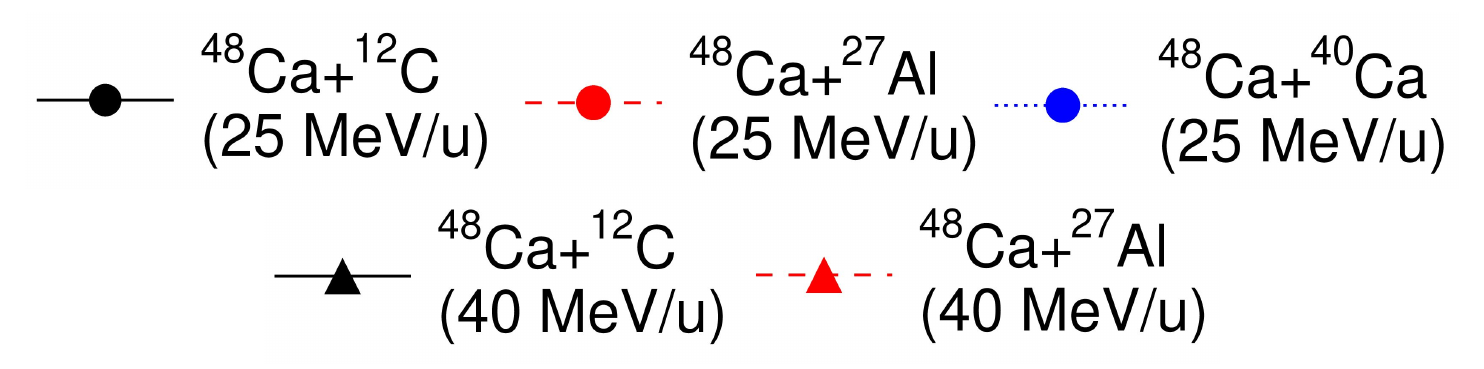}
\includegraphics[width=1.0\columnwidth]{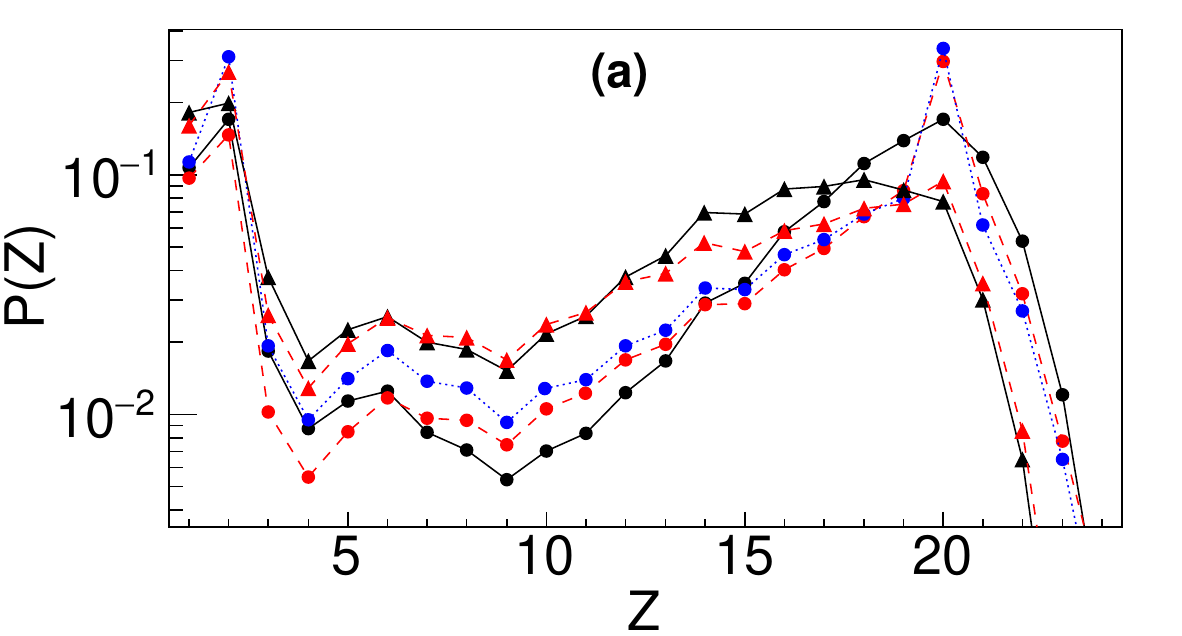}
\includegraphics[width=1.0\columnwidth]{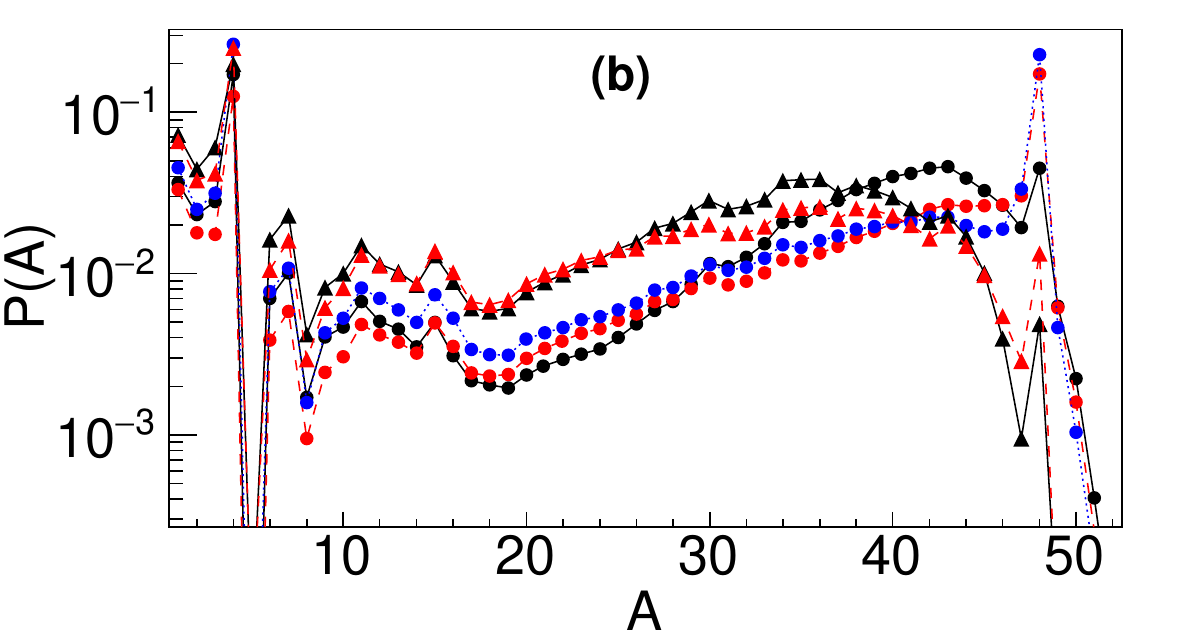}}
\caption{Probability distribution of (a): fragment charge ($Z$) and (b): fragment mass ($A$) for all fragments detected in the FAZIA-PRE experiment. The distributions are normalized to their events. Statistical uncertainties are smaller than the data symbols.}
\label{zanda}
\end{figure}

The characteristics of measured fragments can be better seen, though still in an inclusive way, by plotting the charge and mass distributions for all systems (see Fig. \ref{zanda}). We observe a peak around $Z\sim20$ and $A\sim40$, corresponding to the $QP$ fragments measured in the most peripheral events, while some elastic scattering remains for the low-energy beam reactions which present higher grazing angles (see Table \ref{tab1}). These peaks are clearly emerging (a factor of about 10 more) from the rest of the distributions and give the bright yellow bin in the corresponding correlations in Fig. \ref{zvz}. In other words, from Fig. \ref{zvz} and \ref{zanda}, one can remark that at 25 MeV/u the more symmetric reactions are dominated, as expected, by $QP$ fragments (also very close to the projectile ; $\sim$10 times). The $A$, $Z$ distributions for the more asymmetric case (on $^{12}$C) are broader being associated to a more dispersed primary phase-space mixing various cases of fusion-like events. With increasing bombarding energy, also the $^{27}$Al case changes: the quasi elastic peak is reduced due to the lower grazing angle and the $QP$ sources are more distributed due to larger excitation (see the different shapes of $A$ in the range 30-40 amu at the two energies). We also note, as expected, the abundant production of $Z=1,2$ particles. The gap in the $A$ distribution at $A=5$ is related to the absence of $A=5$ bound nuclei. Also the drop in the mass probability at $A=8$ is due to the breakage of $^{8}$Be into two $^{4}$He.

Systematic uncertainties in the particle identification come from the identification procedure. Larger identification uncertainties result for the low statistics regions of the particle identification spectra. However, their overall contribution is modest and we have verified that the results are not changed much upon reasonable changes in the $Z$, $A$ encoding. Also, the careful calibration procedure on the data ensured an accuracy $\sim$1\% on the kinetic energies of all particles. Moreover, the abundant collected statistics allows to quote small errors on the presented results that often are less than the chosen symbol in the figures.

We now have an overall idea of the range of fragments detected in the FAZIA-PRE experiment. In order to extract more specific information from these fragments, we have used HIPSE event generator to introduce selection rules to be applied to the data.

\section{\label{hipse}HIPSE simulations}

\begin{figure}[!b]
\centering{\includegraphics[width=0.8\columnwidth]{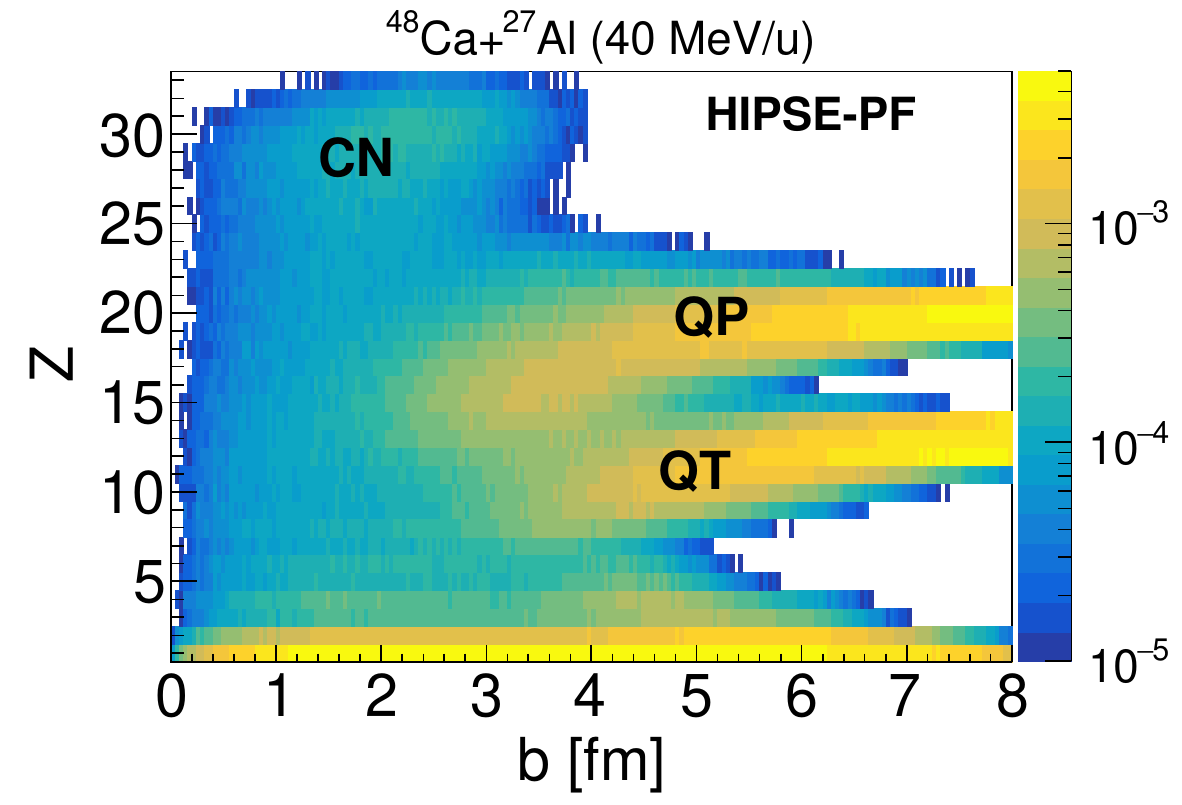}}
\caption{Fragment charge $(Z)$ vs impact parameter $(b)$ for primary fragments (HIPSE-PF) for $^{48}$Ca+$^{27}$Al (40 MeV/u). The correlation shows the full range of $b$ up to the grazing value.}
\label{zvsb}
\end{figure}

\begin{figure}[!b]
\centering{
\includegraphics[width=0.75\columnwidth]{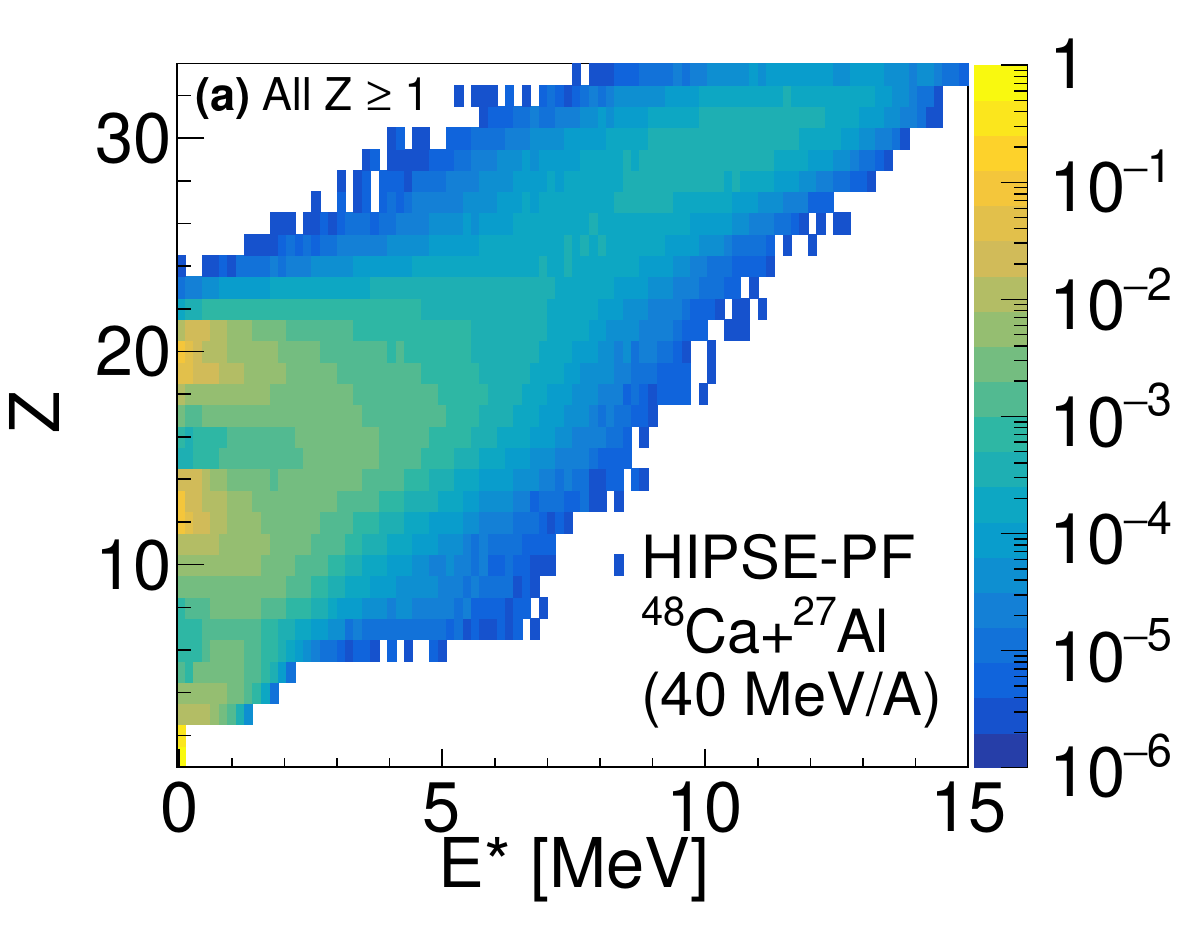}
\includegraphics[width=1.0\columnwidth]{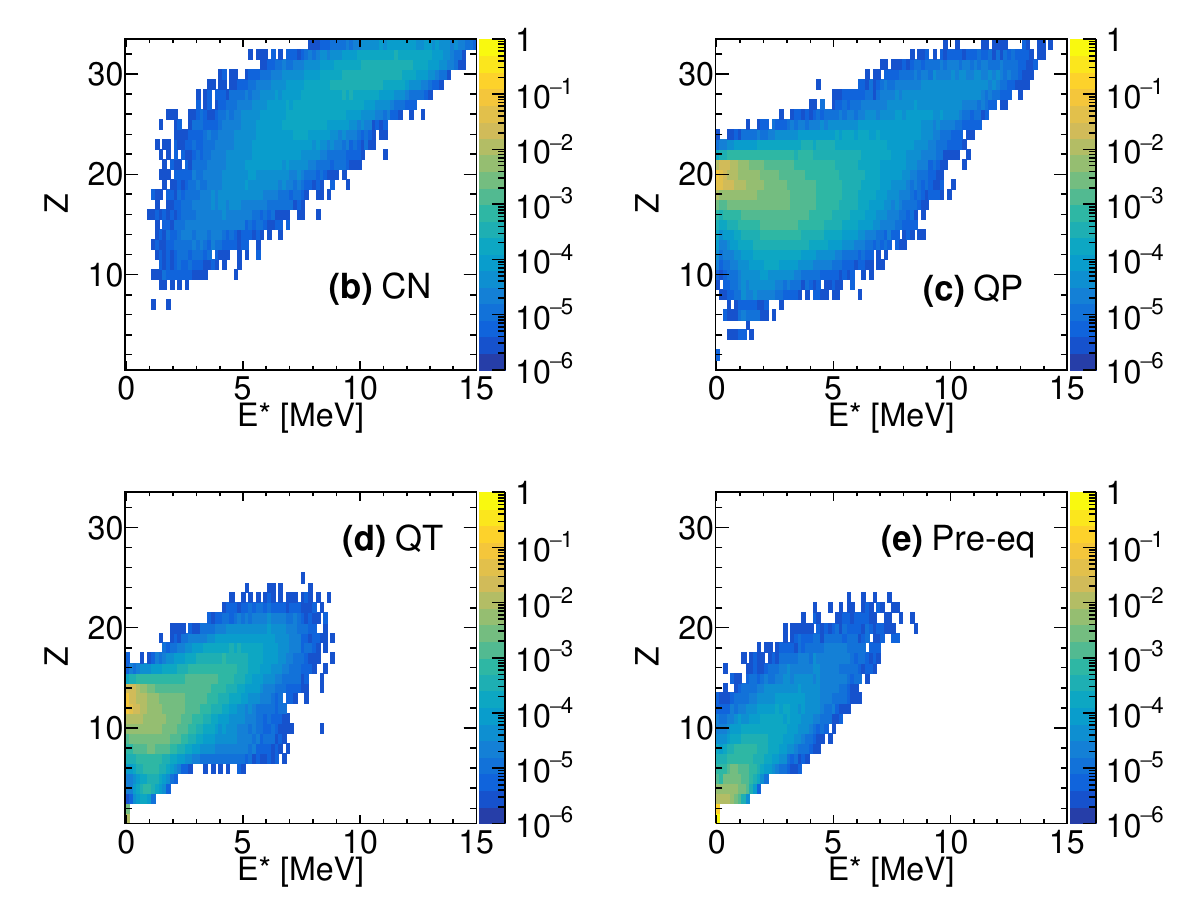}}
\caption{(a): Primary fragment (HIPSE-PF) charge ($Z$) vs excitation energy ($E^{*}$) distribution obtained in reaction $^{48}$Ca+$^{27}$Al (40 MeV/u) system. (b)-(e): Similar to (a) but split into contributions of various emission sources, namely, fusion-like ($CN$), quasi-projectile ($QP$), quasi-target ($QT$) and pre-equilibrium emissions ($Pre-Eq$), respectively. The z-axis for (b)-(e) is same as (a).}\label{zvex}
\end{figure}

The Heavy-Ion Phase Space Exploration (HIPSE) event generator is a semi-phenomenological model, i.e., consisting of both microscopic and macroscopic modelling approaches, which can simulate intermediate energy nuclear reactions at all impact parameters ($b$). It can also reproduce some kind of fast (pre-equilibrium) emissions and has demonstrated to be a relatively fast and reliable reaction simulator from light to heavy systems \cite{PhysRevC.107.044614}, giving, in particular, a good description of $QP$ features for most peripheral collisions \cite{lacroix:in2p3-00620335}. This makes HIPSE, one of the most beneficial models for us to simulate the data with respect to the FAZIA-PRE experiment and extract the quasi-projectile fragments.

Just to give an idea of the events produced by HIPSE simulations, the output for $^{48}$Ca+$^{27}$Al (40 MeV/u) system is presented in this section. For each reaction system, a total of 1 million events were simulated sampled from a triangular distribution of the full $b$ range (0 to the corresponding $b$ grazing). The primary fragments (HIPSE-PF) $Z$ vs $b$ correlation can be seen in Fig. \ref{zvsb}. It is visible that how the various emission sources (marked in the figure) appear with respect to the impact parameter. The excitation energy per nucleon ($E^{*}$) gained by the primary fragments is shown in Fig. \ref{zvex}(a) in correlation with the fragment charge ($Z$). A detailed structure of $E^{*}$ can be observed by splitting these fragments according to the emission sources as labelled in the model output for each event (Fig. \ref{zvex}(b)-(e)). In particular, they are the excited $QP$ and $QT$, fusion like fragments ($CN$) and pre-equilibrium emissions ($Pre$-$Eq$). Clear features appear (see also Fig. \ref{zvsb}). The $QP$, $QT$ populates wide regions in $Z-E^{*}$ corresponding to the impact parameter variation from the grazing collisions (hot spot at low $E^{*}$ and defined charge) to the central collisions. $CN$ yield is increasing with centrality: the sources correspond to the largest and most excited produced fragments (note the $E^{*}$ values even higher than the typical nuclear binding energies). $Pre$-$eq$ fragments are relatively light and their excitation is mostly below 5 MeV. The formation of $QT$, $QP$ is clearly observed from peripheral to semi-central collisions.

\begin{figure}[!b]
\centering{\includegraphics[width=0.91\columnwidth]{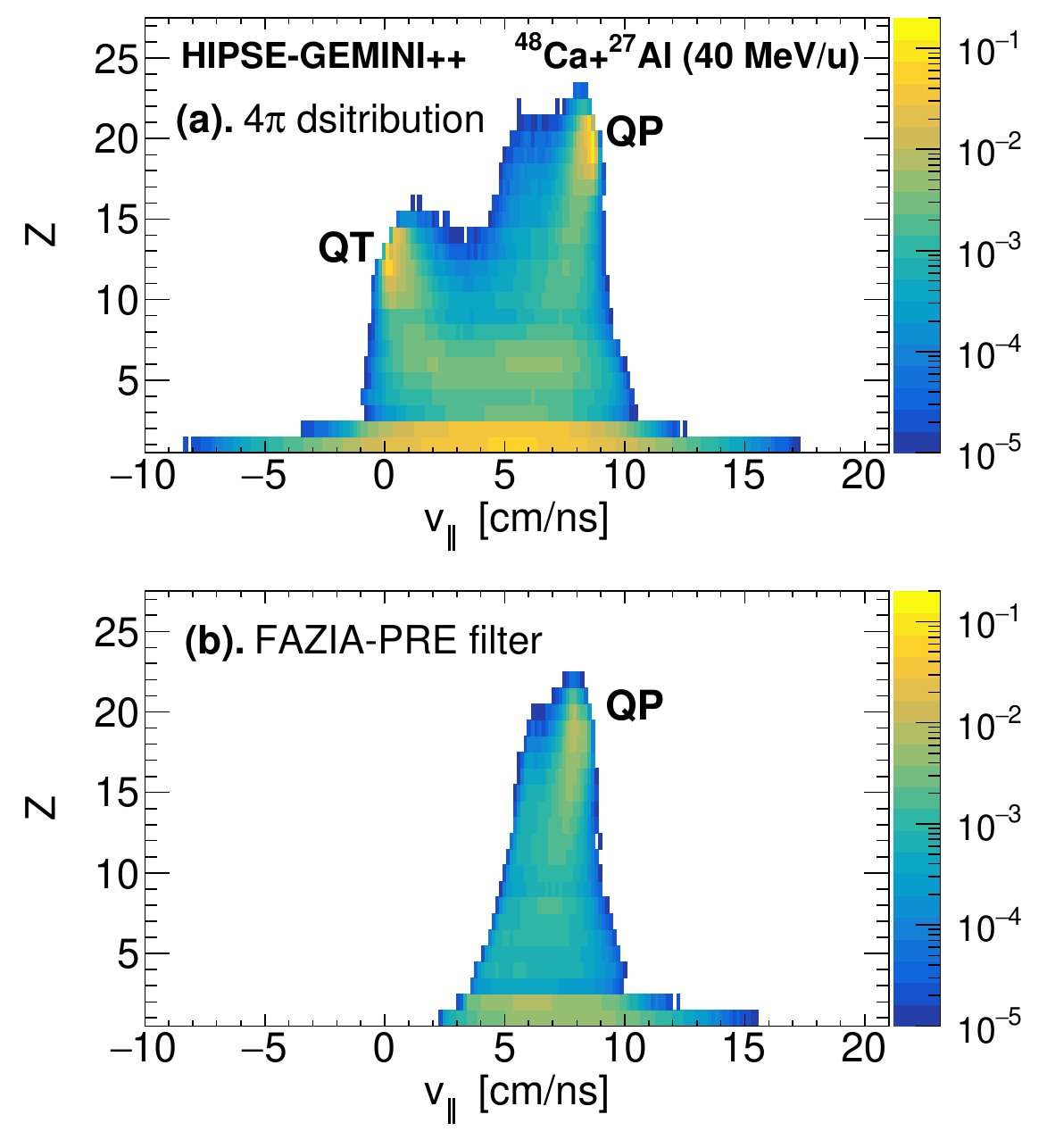}}
\caption{Fragment charge ($Z$) vs longitudinal velocity ($v_{\parallel}$) distribution for $^{48}$Ca+$^{27}$Al (40 MeV/u) system. (a): Full 4$\pi$ HIPSE-GEMINI++ distribution. (b): HIPSE-GEMINI++ simulation after FAZIA-PRE experimental filter. Both data are normalised to their number of events.}
\label{vz_filter}
\end{figure}

\begin{figure}[!b]
\includegraphics[width=1.0\columnwidth]{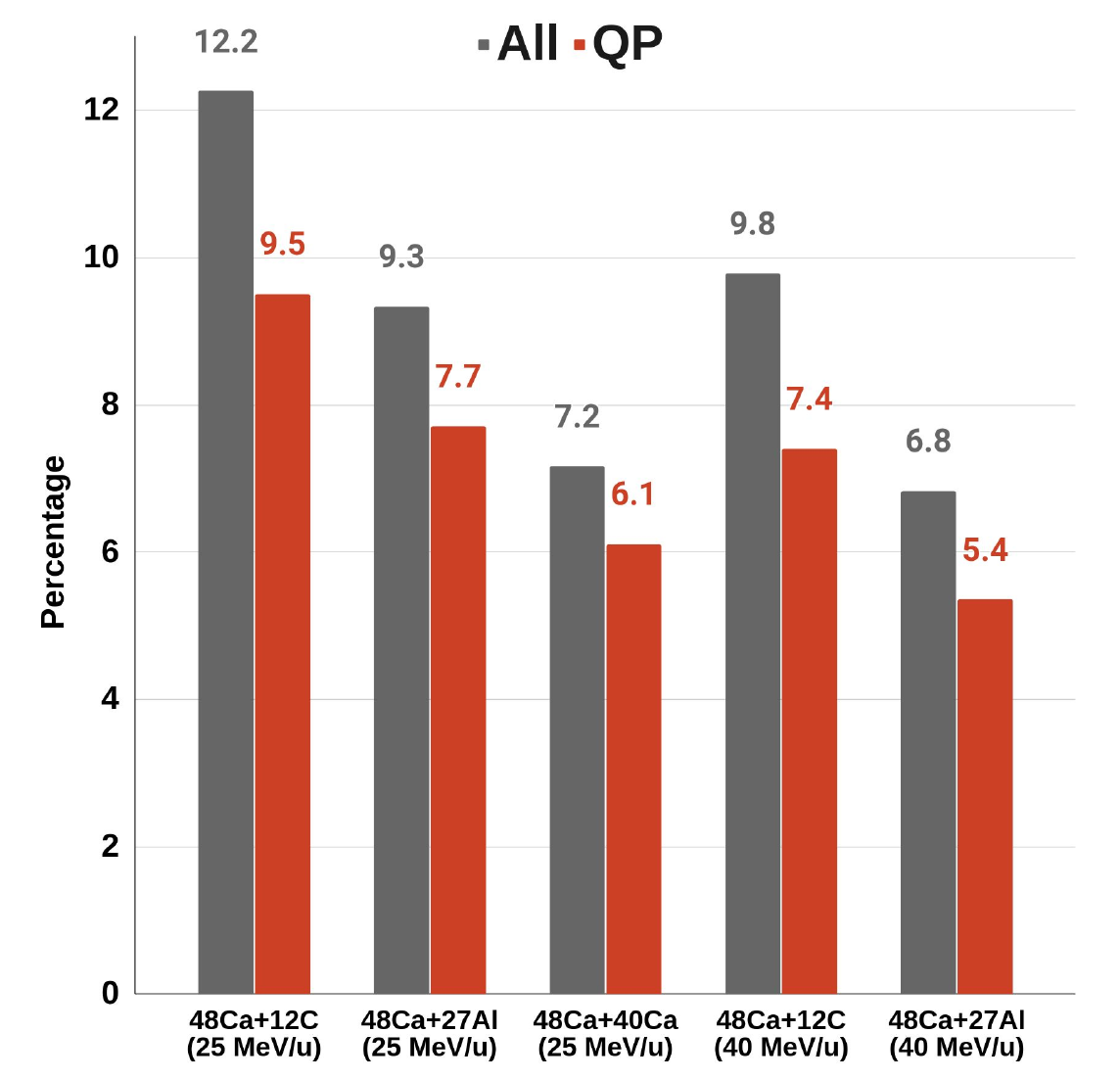}
\caption{Model data: Percentages of detected fragments in the FAZIA-PRE setup, calculated from filtered HIPSE-GEMINI++ simulations. The blue bars correspond to all the charged fragments detected that are coming from various processes and emission sources. The red bars correspond to the detected fragments labelled in HIPSE to be generated from the $QP$ region.}
\label{eff}
\end{figure}

These primary fragment data are passed through de-excitation codes like SIMON \cite{DURAND1992266} or GEMINI++ \cite{Filges2008} to get the final fragments. In this work, we will focus only on the final products from GEMINI++ de-excitation code. Fig. \ref{vz_filter}(a) shows the $Z$ vs $v_{\parallel}$ correlation of full 4$\pi$ distributed HIPSE-GEMINI++ fragments boosted to the laboratory frame. The $QT$ and $QP$ are marked and clearly visible in the figure. The $QT$ region corresponds to $v_{\parallel}\sim0$ as the target is slowly emitted in the lab. On the other hand, the $QP$ region lies near the beam velocity, which for this system is $\sim8.5$ cm/ns. Notwithstanding the good coverage for $QP$ phase-space, the efficiency of the experimental setup is limited and cuts and distortions can be introduced by the detection system. Therefore, the full 4$\pi$ HIPSE-GEMINI++ data are filtered according to the experimental constraints, i.e. setup geometry, energy thresholds, etc. applied in the KaliVeda frameworks \cite{kaliveda}. The output of the filtered simulation is shown in Fig. \ref{vz_filter}(b). Comparing the data before and after applying the experimental filter, it can be observed that as expected, the filtered data well conserves the $QP$ region while removing the $QT$ phase space. We point out the strict similarity of Fig. \ref{vz_filter}(b) with the experimental correlation of the corresponding reaction in Fig. \ref{zvz}(e).

According to the calculations from filtered HIPSE-GEMINI++ simulations, the percentage of detected charged fragment out of the whole 4$\pi$ distribution for the FAZIA-PRE setup is given in Fig. \ref{eff} showing the fragment detection efficiency for all fragments (All, blue bars) and $QP$ fragments (red bars). The efficiency results to be from 7--12\%. However, as expected, according to the model, most ``detected" ejecta (above 70\% overall) are $QP$ fragments; this reinforces the hypothesis that also the measured events are mostly compatible with $QP$ and related emission as we will more precisely discuss in the next section.

\section{\label{comparison}Data comparison and results}

In this section, we will introduce a selection in the experimental data to extract the $QP$ fragments and then compare the experimental results with the filtered HIPSE-GEMINI++ simulations.

We want to limit to the events where the $QP$ fragments do not undergo further break-up or fragmentation; instead, after the decay through particle emission, they are detected as $QP$ remnants. To exclude the lighter fragments and $QP$ break-up channel, we put two conditions. Firstly, we choose only those events in which a single fragment was detected. So, we put the condition on the total charged particle multiplicity to be 1. Secondly, we put a cut on the fragment charge to be $Z>10$, which is greater than half of the projectile $Z$. This selection strongly reduces the probability to have two smaller products from break-up processes, with one of the pair having being missed. According to the model, the remaining spurious events (not-$QP$) with fake $QP$ attribution are well less than 20\% for all investigated reactions. This selection will be referred to as the $QP$-$cut$.

\begin{figure}[!t]
\centering{
\includegraphics[width=0.6\columnwidth]{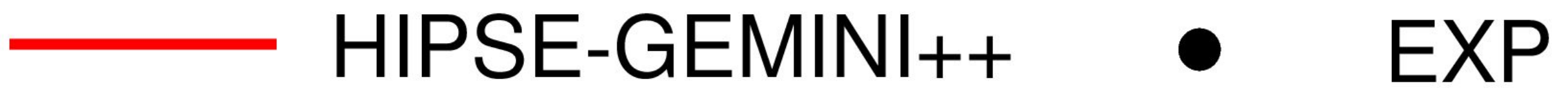}
\includegraphics[width=1.0\columnwidth]{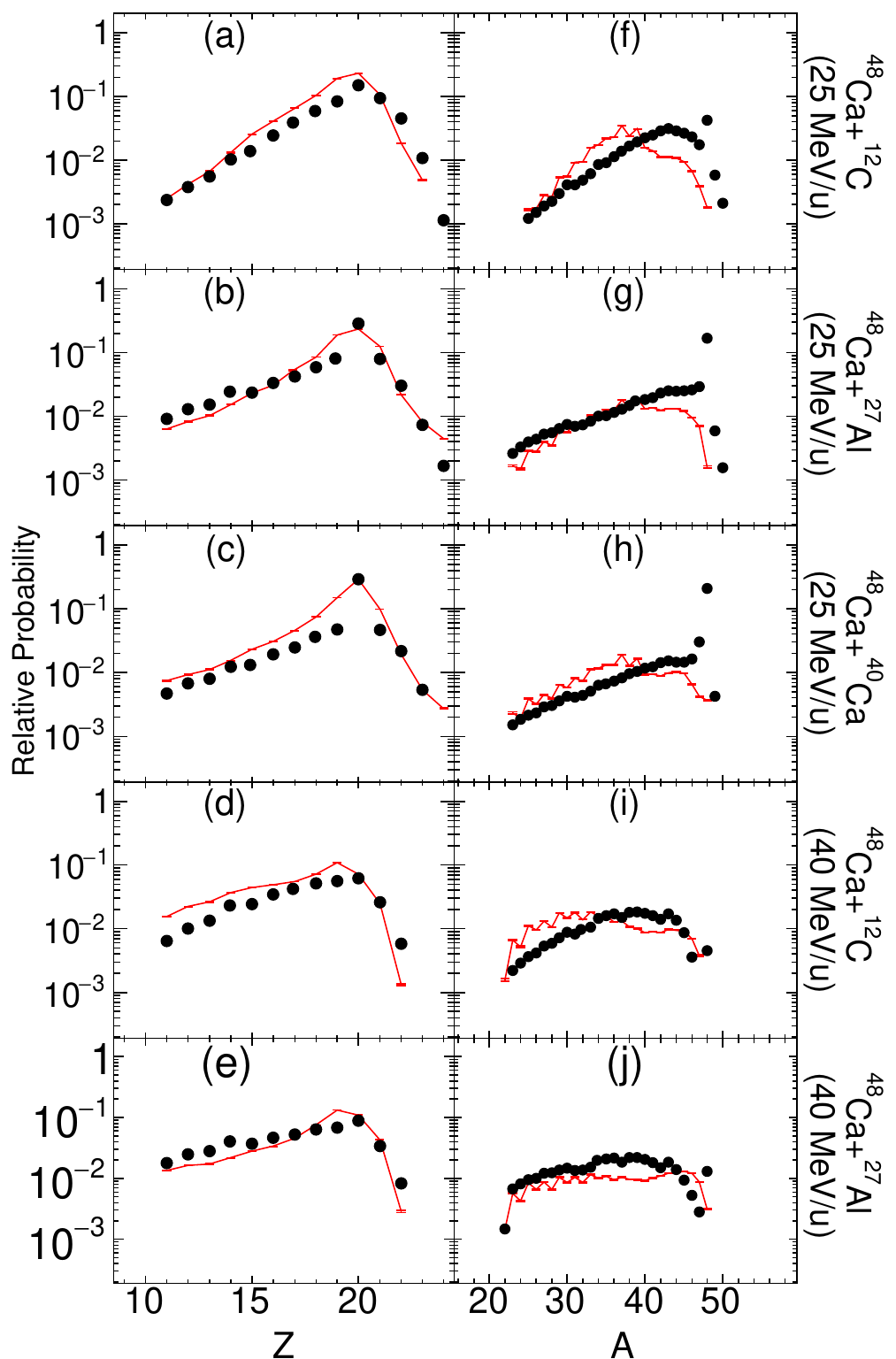}}
\caption{Comparison of the filtered HIPSE-GEMINI++ data with the FAZIA-PRE experimental data (EXP) after applying the $QP$-$cut$. (a)--(e): charge ($Z$) distributions ; (f)--(j): mass ($A$) distributions. Each row represents a reaction system.}
\label{all_comp}
\end{figure}

Now, this $QP$-$cut$ checked within simulated filtered HIPSE-GEMINI++ data is also applied to the experimental data. The direct comparison of HIPSE and the experimental data (EXP) after the $QP$-$cut$ is shown in Fig. \ref{all_comp}. The figure shows the comparison for basic reaction observables, $Z$ (Fig. \ref{all_comp}(a--e)) and $A$ (Fig. \ref{all_comp}(f--j)). Each row represents a reaction system. From the $Z$ distributions, one can infer that HIPSE well reproduces the experimental data. On the other hand, from the $A$ distributions, there is good agreement between the HIPSE and EXP data except for the heaviest masses and more at the lower beam energy. Indeed for these reactions the experimental data contain some residual contribution from very peripheral (almost elastic scattering) events that perhaps the model is not able to predict as a result of very slightly inelastic collisions. Moreover, the model does not include the elastic scattering cross section. The observed mass distribution differences can be explained considering that the projectile is neutron rich, thus the evaporation is contributed more by the emission of free neutrons that immediately change the $A$ distribution. The $Z$ distributions are less modified, so the agreement between model and data remains at a better level than for the nuclear masses.

\begin{figure}[!t]
\centering{
\includegraphics[width=0.6\columnwidth]{legend.pdf}
\includegraphics[width=0.87\columnwidth]{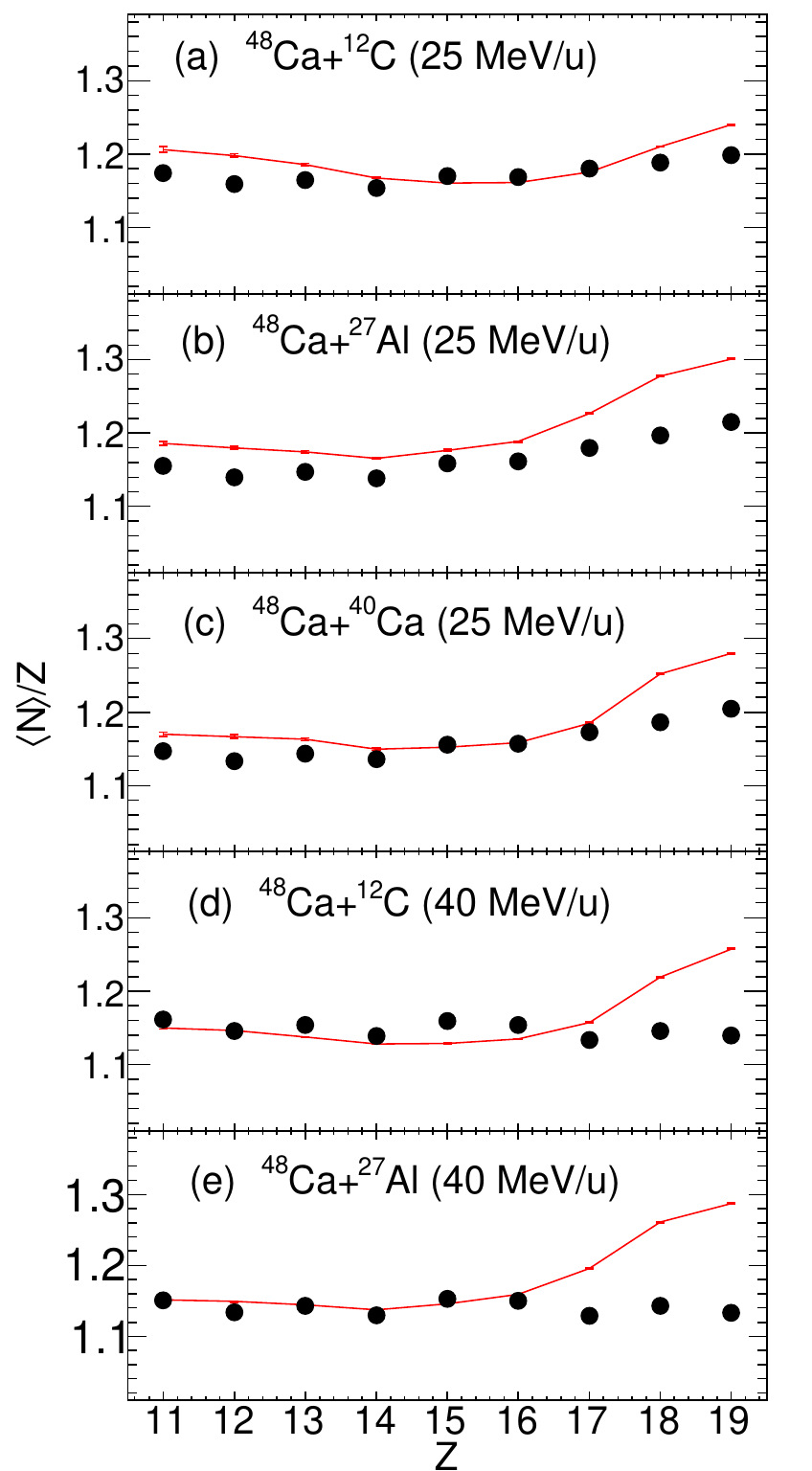}}
\caption{The fragment $\langle N \rangle/Z$ vs $Z$ correlation for all FAZIA-PRE systems to compare HIPSE-GEMINI++ and experimental data (EXP) after applying the $QP$-$cut$.}
\label{nbyz_comp}
\end{figure}

With this caveat in mind and considering the overall observed reasonable agreement between HIPSE and experimental data for basic reaction observables, we also checked the consistency for a more characteristic observable, i.e., the mean fragment isospin ($\langle N \rangle /Z$). The FAZIA detector, with its excellent isotopic resolution allows for refined fragment $N/Z$ studies. Concerning this observable, it is shown in the original HIPSE article \cite{PhysRevC.69.054604} that the calculation gives the final charge to mass ratio of fragments created during the reaction and thus it could be chosen to explore the $N$/$Z$ effects in nuclear collisions in the Fermi-energy range. The fragment $\langle N \rangle /Z$ vs $Z$ is shown in Fig. \ref{nbyz_comp}(a--e) for all FAZIA-PRE systems. The $QP$-$cut$ has been applied to the data presented in the figure. It can be observed from this comparison that HIPSE satisfactorily reproduces the shape of the mean fragment isospin up to $Z\sim16$. Taking into account the previous comment on the limitation of the model towards the most peripheral reaction, the agreement up to $Z\sim16$ is reasonable. Indeed it suggests that the model describes fairly well, the $N/Z$ of fragments coming from rather damped reactions, having final detected charges quite below the projectile value (here $Z$=15--16).

\begin{figure*}[!bt]
\centering{
\includegraphics[width=0.75\textwidth]{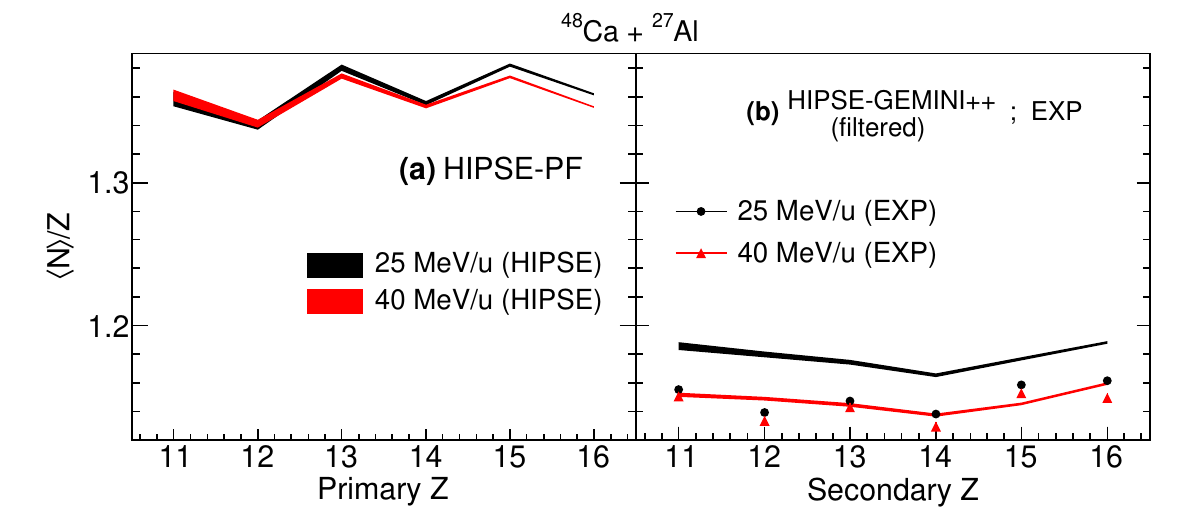}}
\caption{$\langle N\rangle/Z$ vs $Z$ for $^{48}$Ca + $^{27}$Al system after $QP$-$cut$ to observe the effect of changing beam energy. Black band for 25 MeV/u and red band for 40 MeV/u for (a): HIPSE-PF -- primary-fragments; (b): HIPSE-GEMINI++ filtered data + experimental data (black circles for 25 MeV/u and red triangles for 40 MeV/u systems) -- same data from Fig. \ref{nbyz_comp}.}
\label{energydep}
\end{figure*}

Since the entrance channel properties in terms of reacting nuclei result to be rather well described by HIPSE, we now move to study the variation of beam energy, thanks to the two sets of measured data at 25 and 40 MeV/u. We can use the fragment $N/Z$ to observe the entrance channel effects of beam energy. For this, $^{40}$Ca system is not useful because it is only present for 25 MeV/u. As we focus on the $QP$ fragment, the $^{12}$C systems will also not be considered. A detailed study of the $^{12}$C systems from this experiment is given in \cite{piantelli2023}.

In \cite{PhysRevC.55.1900}, it is clearly stated that the additional available energy is mainly removed
by pre-equilibrium emission from the system. The pre-equilibrium neutron multiplicity keeps on increasing with the beam energy. On the other hand, the rate of thermalisation saturates softly after 20 MeV/u \cite{PhysRevC.55.1900}. So, in our case, where the beam energies are 25 and 40 MeV/u, we might expect that the pre-equilibrium emissions are playing a role. This means that if we observe the $QP$ region in the primary fragments, the fragment $N/Z$ should decrease with increasing beam energy. Let's starts with the model predictions. The Fig. \ref{energydep}(a) shows the $\langle N \rangle /Z$ vs $Z$ for HIPSE-PF for $^{48}$Ca+$^{27}$Al system. It can be seen that the $\langle N \rangle /Z$ is generally slightly lower for 40 MeV/u system (especially for the largest remnants). The secondary decays, as expected, then further and strongly reduce the overall fragment $\langle N \rangle /Z$ but preserving the original primary hierarchy even when the FAZIA-PRE experimental constraints are introduced (Fig. \ref{energydep}(b)). Moving to the experimental results, they are drawn with symbols in Fig. \ref{energydep}(b). We see that the data at 40 MeV/u are systematically below those at 25. The effect is smaller than predicted by HIPSE. We are aware that there are multiple effects under consideration that alter the fragment $N/Z$. However, basing on the indication of HIPSE, we can conclude in a qualitative manner that our data are not excluding that the increasing preferential neutron pre-equilibrium emissions with beam energy can partially affect the final $N/Z$ of the $QP$ remnants coming from neutron-rich induced dissipative collisions.

\section{\label{conclusion}Summary and Conclusion}

This work focused on the analysis of the data from the FAZIA-PRE experiment, performed using 6 FAZIA blocks at LNS-INFN (Italy). The measured reaction systems were $^{48}$Ca + $^{12}$C, $^{27}$Al (25, 40 MeV/u) and $^{48}$Ca + $^{40}$Ca (25 MeV/u). The full range of charges and masses of the detected fragments was identified in the experiment. The data for all systems showed high intensity of fragments similar to the projectile as well as around the beam velocities (quasi-projectile fragments).

We used the HIPSE event generator to produce the simulated events in the FAZIA-PRE reaction systems. The HIPSE secondary fragments were obtained in a full 4$\pi$ distribution using the GEMINI++ de-excitation code. This HIPSE-GEMINI++ data were filtered according to the FAZIA-PRE experimental constraints and it was confirmed that the detected fragments are mostly originating from quasi-projectile fragments. Constraints were put following the indications from HIPSE to obtain the single largest quasi-projectile fragment from each collision event. The data from HIPSE and the experiment were then compared after applying these constraints. It was seen that HIPSE gives a good overall reproduction of the experimental data with respect to inclusive fragment observables. Furthermore, HIPSE is able to produce the general trend of the fragment $\langle$N$\rangle$/Z with respect to the experimental data, excluding the $QP$ charges associated to very peripheral collisions, not well described by the model and naturally mixed to more dissipative events in experimental data. The measured average $N/Z$ values of the QP remnants are lower at 40 than at 25 MeV/u beam energy. Although in a qualitative way, the comparison with the model predictions cannot exclude a role of pre-equilibrium emissions that increase with increasing beam energy. Indeed, in combination with the particle evaporation that significantly affects the $N/Z$ of remnants, the slight increasing free neutron fast emissions from the n-rich transient sources before evaporation can also have a role in determining the measured isotopic distributions. This intricate subject can be further explored by comparing the data with other intermediate energy nuclear reaction models like the Antisymmetrized Molecular Dynamics (AMD) model \cite{amd}, Stochastic Mean Field (SMF) model \cite{COLONNA1998449}, etc.

From the results obtained in this work, we can conclude that the HIPSE event generator has proved also in this case to be a fine tool to generate the fragments in nuclear reactions at intermediate energy range and to satisfactorily describe the basic reaction dynamics.

\section*{Acknowledgment}
We would like to thank the accelerator staff of INFN-LNS laboratories for having provided good-quality beams and support during the experiment. This work was partially supported by the POLITA grant: project HARMONIA No. UMO-2014/14/M/ST2/00738 (COPIN-INFN Collaboration) and Jagiellonian University DS research grant (K/DSC/005311/2018).

%
 \bibliographystyle{spphys}
 \bibliography{sahil_fazia}

\begin{thebibliography}{10}
\providecommand{\url}[1]{{#1}}
\providecommand{\urlprefix}{URL }
\expandafter\ifx\csname urlstyle\endcsname\relax
  \providecommand{\doi}[1]{DOI \discretionary{}{}{}#1}\else
  \providecommand{\doi}{DOI \discretionary{}{}{}\begingroup \urlstyle{rm}\Url}\fi

\bibitem{multics}
I.~Iori, et~al., Nuclear Instruments and Methods in Physics Research Section A: Accelerators, Spectrometers, Detectors and Associated Equipment \textbf{325}(3), 458 (1993).
\newblock \doi{10.1016/0168-9002(93)90391-T}

\bibitem{garfield}
F.~Gramegna, et~al., in \emph{IEEE Symposium Conference Record Nuclear Science 2004.}, vol.~2 (2004), vol.~2, pp. 1132--1136 Vol. 2.
\newblock \doi{10.1109/NSSMIC.2004.1462402}

\bibitem{chimera}
S.~Aiello, et~al., Nuclear Physics A \textbf{583}, 461 (1995).
\newblock \doi{10.1016/0375-9474(94)00705-R}.
\newblock Nucleus-Nucleus Collisions

\bibitem{kratta}
J.~Łukasik, et~al., Nuclear Instruments and Methods in Physics Research Section A: Accelerators, Spectrometers, Detectors and Associated Equipment \textbf{709}, 120 (2013).
\newblock \doi{10.1016/j.nima.2013.01.029}

\bibitem{nimrod}
{NIMROD}, {M}ultipurpose charged particle array at {T}exas {A\&M} {U}niversity, \url{https://cyclotron.tamu.edu/nimrod/}

\bibitem{hyra}
N.~Madhavan, et~al., Pramana - J Phys \textbf{75}(2), 317 (2010).
\newblock \doi{10.1007/s12043-010-0119-3}

\bibitem{fazia}
The {FAZIA} {C}ollaboration website, \url{http://fazia.in2p3.fr/}

\bibitem{Bougault2014}
R.~Bougault, et~al., The European Physical Journal A \textbf{50}(2), 47 (2014).
\newblock \doi{10.1140/epja/i2014-14047-4}

\bibitem{PASTORE201742}
G.~Pastore, et~al., Nuclear Instruments and Methods in Physics Research Section A: Accelerators, Spectrometers, Detectors and Associated Equipment \textbf{860}, 42 (2017).
\newblock \doi{10.1016/j.nima.2017.01.048}

\bibitem{piantelli2023}
S.~Piantelli, et~al., Phys. Rev. C \textbf{107}, 044607 (2023).
\newblock \doi{10.1103/PhysRevC.107.044607}

\bibitem{PhysRevC.87.054607}
S.~Barlini, et~al., Phys. Rev. C \textbf{87}, 054607 (2013).
\newblock \doi{10.1103/PhysRevC.87.054607}

\bibitem{PhysRevC.102.044607}
A.~Camaiani, et~al., Phys. Rev. C \textbf{102}, 044607 (2020).
\newblock \doi{10.1103/PhysRevC.102.044607}

\bibitem{PhysRevC.101.034613}
S.~Piantelli, et~al., Phys. Rev. C \textbf{101}, 034613 (2020).
\newblock \doi{10.1103/PhysRevC.101.034613}

\bibitem{PhysRevC.103.014603}
S.~Piantelli, et~al., Phys. Rev. C \textbf{103}, 014603 (2021).
\newblock \doi{10.1103/PhysRevC.103.014603}

\bibitem{PhysRevC.103.014605}
A.~Camaiani, et~al., Phys. Rev. C \textbf{103}, 014605 (2021).
\newblock \doi{10.1103/PhysRevC.103.014605}

\bibitem{PhysRevC.106.024603}
C.~Ciampi, et~al., Phys. Rev. C \textbf{106}, 024603 (2022).
\newblock \doi{10.1103/PhysRevC.106.024603}

\bibitem{Pagano2020}
A.~Pagano, et~al., The European Physical Journal A \textbf{56}(4), 102 (2020).
\newblock \doi{10.1140/epja/s10050-020-00105-z}

\bibitem{epja30.2006}
Dynamics and thermodynamics with nuclear degrees of freedom.
\newblock Edited by P. Chomaz and F. Gulminelli and W. Trautmann and S. Yennello, Eur. Phys. J. A \textbf{30}, thematic issue No. 1, pp. 1-251 (2006)

\bibitem{PhysRevC.79.064614}
E.~Galichet, et~al., Phys. Rev. C \textbf{79}, 064614 (2009).
\newblock \doi{10.1103/PhysRevC.79.064614}

\bibitem{PhysRevC.76.024606}
D.V. Shetty, S.J. Yennello, G.A. Souliotis, Phys. Rev. C \textbf{76}, 024606 (2007).
\newblock \doi{10.1103/PhysRevC.76.024606}

\bibitem{LI2008113}
B.A. Li, L.W. Chen, C.M. Ko, Physics Reports \textbf{464}(4), 113 (2008).
\newblock \doi{10.1016/j.physrep.2008.04.005}

\bibitem{Di_Toro_2010}
M.D. Toro, V.~Baran, M.~Colonna, V.~Greco, Journal of Physics G: Nuclear and Particle Physics \textbf{37}(8), 083101 (2010).
\newblock \doi{10.1088/0954-3899/37/8/083101}

\bibitem{PhysRevLett.84.1120}
F.~Rami, et~al., Phys. Rev. Lett. \textbf{84}, 1120 (2000).
\newblock \doi{10.1103/PhysRevLett.84.1120}

\bibitem{PhysRevC.82.014608}
I.~Lombardo, et~al., Phys. Rev. C \textbf{82}, 014608 (2010).
\newblock \doi{10.1103/PhysRevC.82.014608}

\bibitem{GERACI2004173}
E.~Geraci, et~al., Nuclear Physics A \textbf{732}, 173 (2004).
\newblock \doi{10.1016/j.nuclphysa.2003.11.055}

\bibitem{SOULIOTIS200435}
G.~Souliotis, M.~Veselsky, D.~Shetty, S.~Yennello, Physics Letters B \textbf{588}(1), 35 (2004).
\newblock \doi{10.1016/j.physletb.2004.03.027}

\bibitem{PhysRevLett.92.062701}
M.B. Tsang, et~al., Phys. Rev. Lett. \textbf{92}, 062701 (2004).
\newblock \doi{10.1103/PhysRevLett.92.062701}

\bibitem{PhysRevLett.102.122701}
M.B. Tsang, et~al., Phys. Rev. Lett. \textbf{102}, 122701 (2009).
\newblock \doi{10.1103/PhysRevLett.102.122701}

\bibitem{PhysRevC.82.051603}
Z.Y. Sun, et~al., Phys. Rev. C \textbf{82}, 051603 (2010).
\newblock \doi{10.1103/PhysRevC.82.051603}

\bibitem{PhysRevC.62.041605}
M.~Veselsky, et~al., Phys. Rev. C \textbf{62}, 041605 (2000).
\newblock \doi{10.1103/PhysRevC.62.041605}

\bibitem{PhysRevC.86.014610}
E.~De~Filippo, et~al., Phys. Rev. C \textbf{86}, 014610 (2012).
\newblock \doi{10.1103/PhysRevC.86.014610}

\bibitem{BARAN2005335}
V.~Baran, M.~Colonna, V.~Greco, M.~{Di Toro}, Physics Reports \textbf{410}(5), 335 (2005).
\newblock \doi{10.1016/j.physrep.2004.12.004}

\bibitem{PhysRevLett.94.032701}
L.W. Chen, C.M. Ko, B.A. Li, Phys. Rev. Lett. \textbf{94}, 032701 (2005).
\newblock \doi{10.1103/PhysRevLett.94.032701}

\bibitem{PhysRevC.76.034603}
T.X. Liu, et~al., Phys. Rev. C \textbf{76}, 034603 (2007).
\newblock \doi{10.1103/PhysRevC.76.034603}

\bibitem{23}
B.A. Li, W.U. Schr{\"o}der, \emph{Isospin Physics in Heavy-Ion Collisions at Intermediate Energies} (Nova Science, New York, 2001)

\bibitem{PhysRevC.68.024605}
G.A. Souliotis, et~al., Phys. Rev. C \textbf{68}, 024605 (2003).
\newblock \doi{10.1103/PhysRevC.68.024605}

\bibitem{COLONNA2008454c}
M.~Colonna, J.~Rizzo, P.~Chomaz, M.~{Di Toro}, Nuclear Physics A \textbf{805}(1), 454c (2008).
\newblock \doi{10.1016/j.nuclphysa.2008.02.266}.
\newblock INPC 2007

\bibitem{PhysRevC.80.014322}
L.W. Chen, B.J. Cai, C.M. Ko, B.A. Li, C.~Shen, J.~Xu, Phys. Rev. C \textbf{80}, 014322 (2009).
\newblock \doi{10.1103/PhysRevC.80.014322}

\bibitem{PhysRevC.79.064615}
E.~Galichet, M.~Colonna, B.~Borderie, M.F. Rivet, Phys. Rev. C \textbf{79}, 064615 (2009).
\newblock \doi{10.1103/PhysRevC.79.064615}

\bibitem{fable2023study}
Q.~Fable, et~al.
\newblock arxiv:2312.01763 \textbf{[nucl-ex]} (2023).
\newblock \doi{10.48550/arXiv.2312.01763}

\bibitem{PhysRevC.37.1783}
R.T. de~Souza, W.U. Schr\"oder, J.R. Huizenga, R.~Planeta, K.~Kwiatkowski, V.E. Viola, H.~Breuer, Phys. Rev. C \textbf{37}, 1783 (1988).
\newblock \doi{10.1103/PhysRevC.37.1783}

\bibitem{PhysRevC.38.195}
R.~Planeta, et~al., Phys. Rev. C \textbf{38}, 195 (1988).
\newblock \doi{10.1103/PhysRevC.38.195}

\bibitem{PhysRevC.74.051602}
D.~Th\'eriault, et~al., Phys. Rev. C \textbf{74}, 051602 (2006).
\newblock \doi{10.1103/PhysRevC.74.051602}

\bibitem{PhysRevC.81.034603}
A.B. McIntosh, et~al., Phys. Rev. C \textbf{81}, 034603 (2010).
\newblock \doi{10.1103/PhysRevC.81.034603}

\bibitem{PhysRevC.86.021603}
S.~Hudan, et~al., Phys. Rev. C \textbf{86}, 021603 (2012).
\newblock \doi{10.1103/PhysRevC.86.021603}

\bibitem{PhysRevC.106.024605}
Q.~Fable, et~al., Phys. Rev. C \textbf{106}, 024605 (2022).
\newblock \doi{10.1103/PhysRevC.106.024605}

\bibitem{Ciampi_2023}
C.~Ciampi, et~al., Journal of Physics: Conference Series \textbf{2586}(1), 012039 (2023).
\newblock \doi{10.1088/1742-6596/2586/1/012039}

\bibitem{PhysRevC.107.014604}
Q.~Fable, et~al., Phys. Rev. C \textbf{107}, 014604 (2023).
\newblock \doi{10.1103/PhysRevC.107.014604}

\bibitem{PhysRevC.108.054611}
C.~Ciampi, et~al., Phys. Rev. C \textbf{108}, 054611 (2023).
\newblock \doi{10.1103/PhysRevC.108.054611}

\bibitem{PhysRevC.69.054604}
D.~Lacroix, A.~Van~Lauwe, D.~Durand, Phys. Rev. C \textbf{69}, 054604 (2004).
\newblock \doi{10.1103/PhysRevC.69.054604}

\bibitem{lacroix:in2p3-00023917}
D.~Lacroix, A.~Van~Lauwe, D.D. Durand, in \emph{{International Workshop on Multifragmentation and Related Topics (IWM2003)}}, ed. by L.~Wieleczko ({Infn Catania}, CAEN, France, 2003), pp. 58--61

\bibitem{PhysRevC.55.1900}
L.~Lassen, P.~von Neumann-Cosel, A.~Oberstedt, G.~Schrieder, Phys. Rev. C \textbf{55}, 1900 (1997).
\newblock \doi{10.1103/PhysRevC.55.1900}

\bibitem{PhysRevC.107.044614}
C.~Frosin, et~al., Phys. Rev. C \textbf{107}, 044614 (2023).
\newblock \doi{10.1103/PhysRevC.107.044614}

\bibitem{lacroix:in2p3-00620335}
D.~Lacroix, {Quantum nuclear many-body dynamics and related aspects}.
\newblock Research report, {CNRS} (2011).
\newblock M{\'e}moire de HDR soutenu le 16 d{\'e}cembre 2010, Universit{\'e} de Caen-Basse Normandie

\bibitem{DURAND1992266}
D.~Durand, Nuclear Physics A \textbf{541}(2), 266 (1992).
\newblock \doi{10.1016/0375-9474(92)90097-4}

\bibitem{Filges2008}
R.J. Charity, Joint ICTP-IAEA advanced workshop on model codes for spallation reactions, Trieste, Italy, 2008, IAEA, Report INDC(NDC)-0530. p. 139 (2008)

\bibitem{kaliveda}
Kali{V}eda {T}oolkit, \url{https://kaliveda.in2p3.fr/}

\bibitem{amd}
Y.~Kanada-En'yo, M.~Kimura, A.~Ono, Progress of Theoretical and Experimental Physics \textbf{2012}(1) (2012).
\newblock \doi{10.1093/ptep/pts001}.
\newblock 01A202

\bibitem{COLONNA1998449}
M.~Colonna, et~al., Nuclear Physics A \textbf{642}(3), 449 (1998).
\newblock \doi{10.1016/S0375-9474(98)00542-9}

\end{thebibliography}

\end{document}